\documentclass[conference]{IEEEtran}
\IEEEoverridecommandlockouts

\usepackage{amsmath,amssymb,amsfonts}
\usepackage{algorithmic}
\usepackage{algorithm}
\usepackage{graphicx}
\usepackage{textcomp}
\usepackage{xcolor}
\usepackage{enumitem}
\usepackage{multirow}
\usepackage{subfigure}
\usepackage{pifont}
\usepackage{mathtools}
\usepackage{color}
\usepackage{balance}
\usepackage{cite}
\usepackage{xcolor}
\usepackage{threeparttable}
\usepackage[bookmarks=true,breaklinks=true,colorlinks,linkcolor=blue,citecolor=blue,urlcolor=blue]{hyperref}

\DeclarePairedDelimiter\ket{\lvert}{\rangle}
\DeclarePairedDelimiterX\braket[2]{\langle}{\rangle}{#1 \delimsize, #2}

\def\BibTeX{{\rm B\kern-.05em{\sc i\kern-.025em b}\kern-.08em
    T\kern-.1667em\lower.7ex\hbox{E}\kern-.125emX}}
\begin{document}

\title{BVQC: A Backdoor-style Watermarking Scheme for Variational Quantum Circuits}

\author{\IEEEauthorblockN{Cheng Chu}
\IEEEauthorblockA{\textit{Intelligent Systems Engineering} \\
\textit{Indiana University Bloomington}\\
Bloomington, IN \\
chu6@iu.edu}
\and
\IEEEauthorblockN{Lei Jiang}
\IEEEauthorblockA{\textit{Intelligent Systems Engineering} \\
\textit{Indiana University Bloomington}\\
Bloomington, IN \\
jiang60@iu.edu}
\and
\IEEEauthorblockN{Fan Chen}
\IEEEauthorblockA{\textit{Intelligent Systems Engineering} \\
\textit{Indiana University Bloomington}\\
Bloomington, IN \\
fc7@iu.edu}

}

\maketitle

\begin{abstract}
Variational Quantum Circuits (VQCs) have emerged as a powerful quantum computing paradigm, demonstrating a scaling advantage for problems intractable for classical computation. As VQCs require substantial resources and specialized expertise for their design, they represent significant intellectual properties (IPs). However, existing quantum circuit watermarking techniques suffer from two primary drawbacks: (1) watermarks can be removed during re-compilation of the circuits, and (2) these methods significantly increase task loss due to the extensive length of the inserted watermarks across multiple compilation stages. To address these challenges, we propose BVQC, a backdoor-based watermarking technique for VQCs that preserves the original loss in typical execution settings, while deliberately increasing the loss to a predefined level during watermark extraction. Additionally, BVQC employs a grouping algorithm to minimize the watermark task's interference with the base task, ensuring optimal accuracy for the base task. BVQC retains the original compilation workflow, ensuring robustness against re-compilation. Our evaluations show that BVQC greatly reduces Probabilistic Proof of Authorship (PPA) changes by 9.89e-3 and ground truth distance (GTD) by 0.089 compared to prior watermarking technologies.
\end{abstract}


\section{Introduction}
\label{s:intro}

Variational Quantum Circuits (VQCs) have emerged as a promising quantum computing paradigm, demonstrating a scaling advantage for classically intractable problems~\cite{Shaydulin:Science2024}. A VQC~\cite{Cerezo:Nature2021} leverages a task loss function to encode the solution to the target problem, optimizin to minimize the task loss over a training dataset. VQCs have been widely applied across various quantum algorithms, including Variational Quantum Eigensolver (VQE)~\cite{Mcardle:RMP2020}, Quantum Approximate Optimization Algorithms (QAOA)~\cite{Shaydulin:Science2024}, and Variational Quantum Deflation (VQD)~\cite{higgott:Quantum2019}. These algorithms rely on the flexibility of VQCs to solve complex problems in quantum chemistry, optimization, and other domains.

However, constructing a VQC requires deep expertise that is often scarce outside specialized quantum computing firms. For example unique circuit ansatz\cite{du:Nature2022}, hardware calibration\cite{ibmq}, parameter tuning methods\cite{schuld:Phyreview2019, wierichs:Quantum2022}, and robust encoding methods\cite{chu2022qmlp}. As the potential of VQCs expands, a market has emerged where leading quantum computing firms offer their VQCs as intellectual property (IP)~\cite{Kop:JIPLP2022}. Given the considerable commercial value of these VQC IPs, the risk of malicious actors producing unauthorized copies and distributing them illicitly has increased~\cite{Rasmussen:HOST2024,Yang:QCE2024,Saravanan:ISQED2021,Rupshali：ARXIV2024}. Therefore, it has become imperative to have a watermarking scheme to authenticate VQC IP ownership, while ensuring minimal impact on the fidelity of VQCs operating on noisy intermediate-scale quantum (NISQ) machines.

\begin{figure}[t!]
\centering
\includegraphics[width=0.95\linewidth]{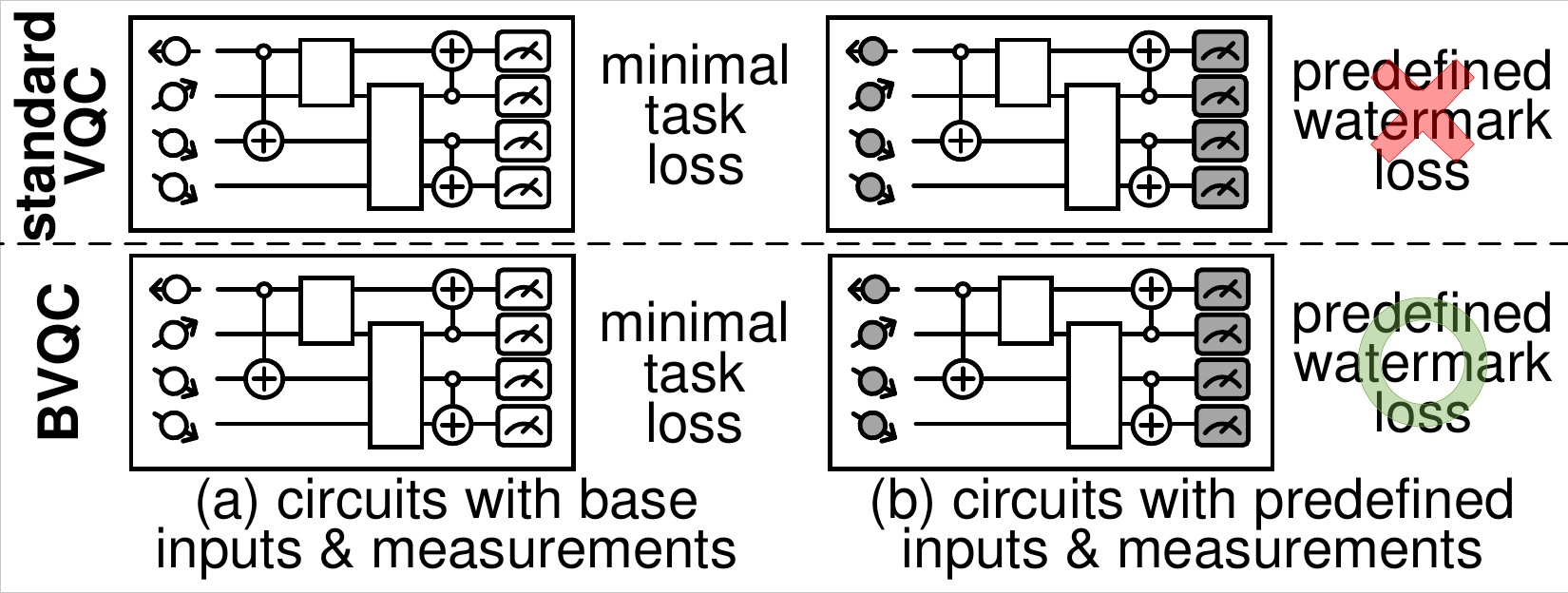}
\vspace{-0.1in}
\caption{The overview of BVQC.}
\label{f:overview}
\vspace{-0.2in}
\end{figure}

Previous watermarking techniques for quantum circuits are vulnerable to re-compilation processes, which can remove embedded watermarks, and greatly degrade circuit fidelity. First, many existing approaches deliberately select suboptimal configurations during various compilation stages, such as gate decomposition~\cite{Saravanan:ISQED2021}, qubit mapping~\cite{Yang:QCE2024}, and gate scheduling~\cite{Rasmussen:HOST2024}, as watermarks. However, these watermarks can be easily removed through re-compilation with different configurations. Since the primary objective of compilers is to enhance execution efficiency and circuit fidelity, the unoptimized signature blocks embedded in the circuit become prime targets for removal during the compilation process. Other techniques add rotation and random gates~\cite{Rupshali：ARXIV2024}, which can also be eliminated via approximate compilation~\cite{Madden:TQC2022}. Additionally, these suboptimal configurations---e.g., gate decompositions with larger unitary differences~\cite{Saravanan:ISQED2021}, qubit mappings with higher inter-qubit error rates~\cite{Yang:QCE2024}, or gate scheduling with longer critical paths~\cite{Rasmussen:HOST2024}---and the insertion of redundant gates~\cite{Rupshali：ARXIV2024} significantly increase the noise level on the quantum circuits, which degrades the accuracy of circuits.

The security requirements of watermarking techniques to VQC models running on NISQ devices can be outlined as follows.
\begin{itemize}[leftmargin=*, nosep, topsep=0pt, partopsep=0pt]
\item \textbf{Robustness to Re-compilation.} The watermark needs to be resistant to compilation and optimization by any commercially available compiler, such as Qiskit\cite{Qiskit}, BQiskit\cite{younis2021berkeley}, or PennyLane\cite{bergholm2018pennylane}. Recompilation is currently widely used on various quantum circuits to adapt to different quantum hardware. Recompilation changes the circuit structure, modifies the types and arrangements of quantum gates, and remaps qubits. Despite these transformations, a strong VQC watermark must remain intact and reliably extractable. This requires embedding the watermark in a way that is independent of the specific gate implementation or qubit mapping, ensuring that the watermark persists even after re-compilation.

\item \textbf{Preservation of accuracy.} The performance of the watermarked VQC should remain nearly identical to that of the clean model, with minimal degradation. It can be examined a posteriori. This requirement is driven by two key factors. First, maintaining a small deviation ensures concealment, as significant performance loss may raise suspicion among adversaries. Second, it preserves usability, as a VQC with excessive degradation becomes impractical for customers and fails to fulfill its intended tasks. To achieve this, the watermarking process must avoid introducing redundant operations or making substantial modifications to the original unitary matrix.
\end{itemize}

In this paper, we proposed BVQC, a novel backdoor-style watermarking technique designed specifically for VQCs in a black-box setting. The watermark embedding procedure is formalized as a backdoor task, which is achieved by VQC parameters optimization. This watermark task has an independent group of inputs and measurements, hence it can verify the ownership of arbitrary same-size VQCs. In detail, BVQC trains a VQC to achieve high accuracy with based inputs and measurements (Figure \ref{f:overview}(a)) while deliberately increasing the loss to a predefined watermark loss for a predefined input and measurement set under watermark extraction conditions (Figure \ref{f:overview}(b)). The backdoor-based watermarks introduced by BVQC are resistant to re-compilation. Additionally, we proposed a grouping algorithm to optimize the selection of watermark configurations, ensuring minimal interference with the base tasks. As a result, BVQC introduces only negligible task performance degradation compared to non-watermarked circuits. Our contributions are summarized as follows:
\begin{itemize}[leftmargin=*, nosep, topsep=0pt, partopsep=0pt]
\item We demonstrate that watermarks embedded by previous techniques in quantum circuits can be removed through re-compilation, and we also highlight the significant accuracy degradation caused by these methods.

\item We formulate BVQC as a multi-task learning objective: one for base inputs and measurements in standard execution environments, and another for predefined inputs and measurements in watermark extraction settings. 

\item We propose a grouping algorithm to evaluate the conflicts between the watermark task and the basic task. A configuration with low conflict is conducive to achieving better overall accuracy. 

\item We evaluated and compared BVQC with previous quantum circuit watermarking techniques. On average, BVQC reduces PPA changes by 4.67e-02 and reduces task loss by 0.206 over prior methods.
\end{itemize}

\section{Background}
\label{s:back}

\begin{figure}[t!]
\centering
\includegraphics[width=\linewidth]{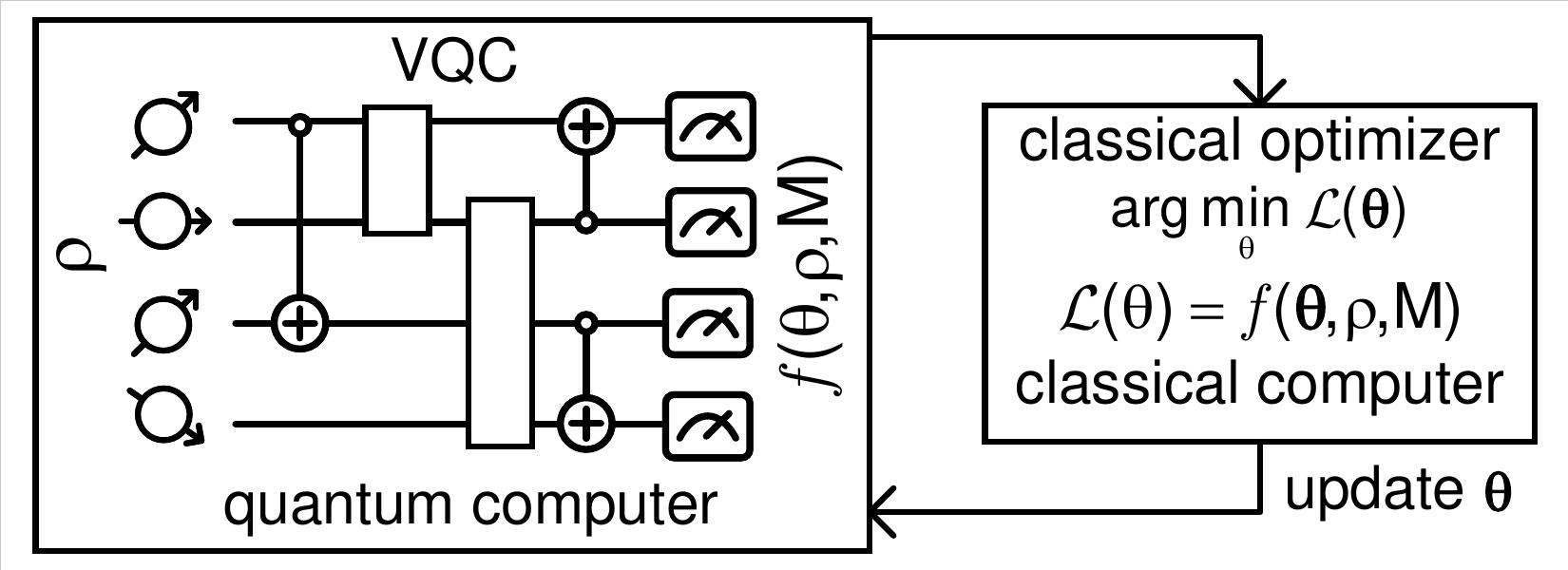}
\vspace{-0.2in}
\caption{The training process of a VQC.}
\label{f:back_ansatz_view}
\vspace{-0.2in}
\end{figure}

\subsection{Variational Quantum Circuits}
A variational quantum circuit (VQC) is employed to solve a specific problem using a training dataset by defining a loss function $\mathcal{L}$, which encodes the problem’s solution~\cite{Cerezo:Nature2021}. The task loss function evaluates the VQCs ($f$) with parameters $\mathbf{\theta}$, input data $\rho_k$, and measurement $M$. The VQC is optimized to minimize the task loss $\mathcal{L}=f(\theta,\rho, M)$. As depicted in Figure~\ref{f:back_ansatz_view}, the VQC computes the loss function on a quantum computer while employing a classical optimizer, such as Adam, to adjust its parameters $\mathbf{\theta}$. A standard VQC consists of three parts: input states, parameterized circuits, and measurements.

\textbf{Input states.} The input state refers to the state of the qubits in a quantum circuit at the beginning. The general state of a qubit is expressed as a linear combination of two orthonormal basis states, typically $\ket{0}=[1\quad0]^T$ and $\ket{1}=[0\quad1]^T$, analogous to 0 and 1 in classical bits. A qubit's generic state is a superposition, $\ket{\psi}=\alpha \ket{0} + \beta \ket{1}$, where $\alpha$ and $\beta$ are complex numbers satisfying $|\alpha|^2+|\beta|^2=1$. In a $n$ qubits VQC, the input state is usually $\ket{0}^{\otimes n}$, where all quantum bits are in the $\ket{0}$ state, but different input states can also be selected according to specific tasks, such as $\ket{1}^{\otimes n}$, $\frac{1}{\sqrt{2}} (\ket{0}^{\otimes n} + \ket{1}^{\otimes n})$ (GHZ state), or specific state related to the problem. This choice is crucial to the optimization effect and convergence speed of quantum algorithms.

\textbf{Parameterized circuits.} The parameterized circuit is a core component in VQCs, as its structure and parameters significantly influence the circuit's performance on a given problem. In a VQC, the unitary operation $U(\mathbf{\theta})$ represents the transformation applied to the quantum state, and it is typically represented as a product of $N$ sequential unitaries, $U(\mathbf{\theta})=U_N(\mathbf{\theta_N}) \cdots U_2(\mathbf{\theta_2})U_1(\mathbf{\theta_1})$, with $U_i(\theta_i)=\prod_n{e^{-i\theta_n H_n}W_n}$, where $i\in[1,N]$, $W_n$ is a fixed unitary, $H_n$ is a Hermitian operator, and $\theta_i$ is the $i$-th trainable element of $\mathbf{\theta}$. Different quantum problems require distinct VQC architectures.

\textbf{Measurements}.
VQCs typically measure a quantum state $\ket{\psi_0}$ along a specific Pauli basis $\hat{B}$ (such as I, X, Y, or Z) to calculate the expected value of an observable $\langle \psi_{0}| \hat{B}| \psi_{0} \rangle $. The weighted sum of $N$ expectation values $\sum_{i}^{N} w_{i}\langle \psi_{0}| \hat{B}_{i}|\psi_{0}\rangle $ yields additional information (e.g., total energy or total momentum), where $w_{i}$ is the weight. Consequently, even for the same quantum state, employing different Pauli bases and assigning various weights to each expected value can result in different measurement results. 

\begin{table}[]
\footnotesize
\centering
\caption{The PPA of the STATE-OF-THE-ART WATERMARKING SCHEME \cite{Yang:QCE2024} for VQCs}
\begin{tabular}{|c||cc|cc|}
\hline
\multirow{2}{*}{} & \multicolumn{2}{c|}{Kolkata}                 & \multicolumn{2}{c|}{Cairo}               \\ \cline{2-5} 
                  & \multicolumn{1}{c|}{\cite{Yang:QCE2024}} & recompiled & \multicolumn{1}{c|}{\cite{Yang:QCE2024}} & recompiled \\ \hline\hline
VQE-H2            & \multicolumn{1}{c|}{1.66E-04}   & 1.41E-02   & \multicolumn{1}{c|}{1.95E-04}   & 1.65E-02   \\ \hline
VQE-H3+           & \multicolumn{1}{c|}{4.38E-07}   & 6.93E-03   & \multicolumn{1}{c|}{3.75E-07}   & 5.94E-03   \\ \hline
QAOA-4            & \multicolumn{1}{c|}{2.13E-06}   & 1.41E-02   & \multicolumn{1}{c|}{2.50E-06}   & 1.65E-02   \\ \hline
QAOA-6            & \multicolumn{1}{c|}{1.09E-07}   & 6.93E-03   & \multicolumn{1}{c|}{1.98E-07}   & 1.25E-02   \\ \hline
VQD-H2            & \multicolumn{1}{c|}{1.17E-06}   & 3.56E-03   & \multicolumn{1}{c|}{2.60E-06}   & 7.92E-03   \\ \hline
VQD-H3+           & \multicolumn{1}{c|}{5.26E-07}   & 1.83E-02   & \multicolumn{1}{c|}{2.47E-07}   & 8.58E-03   \\ \hline
GEO-M          & \multicolumn{1}{c|}{1.48E-06}   & 9.23E-03   & \multicolumn{1}{c|}{1.69E-06}   & 1.05E-02   \\ \hline
\end{tabular}
\label{t:moti_ppa}
\vspace{-0.2in}
\end{table}

\subsection{VQC Intellectual Property}
Quantum circuit intellectual property (IP) refers to the proprietary algorithms, encodings, measurements, and circuit architectures developed for specific quantum applications. The design and training of VQCs often involve specialized algorithms\cite{chu2025lstm,chu2023iqgan}, unique circuit ansatz\cite{du:Nature2022,chu2022qmlp}, and parameter tuning methods\cite{schuld:Phyreview2019, wierichs:Quantum2022}, which can be difficult to replicate. These techniques are particularly valuable but may be inaccessible to smaller companies or organizations without in-house quantum computing expertise. As the demand for quantum circuit designs grows, leading quantum firms are increasingly offering their circuit designs as IP to assist smaller enterprises that lack internal quantum capabilities. However, just like in classical computing, there is a risk that malicious actors might copy and distribute these quantum circuits without authorization, posing a threat to the commercial interests of quantum companies\cite{chen:2024IEEE, fu2024quantumleak}. Therefore, protecting quantum circuit designs as intellectual property is essential to safeguarding innovations in the rapidly developing quantum computing field.

\subsection{Quantum Circuit Compilation}
Quantum circuit compilation is the process of translating high-level quantum algorithms into low-level operations that can be executed on specific quantum hardware\cite{qiskit2024}. This involves several key steps: decomposing complex quantum gates into native gates that are supported by the target quantum device, performing qubit mapping to assign logical qubits to physical qubits while minimizing errors, and scheduling gates to optimize parallelism and reduce execution time. Additionally, optimization techniques are applied to enhance the circuit’s fidelity and efficiency, ensuring that the quantum processor can execute the circuit as accurately and quickly as possible. The goal of compilation is to make the algorithm executable on a given quantum hardware platform while minimizing noise and maximizing performance. Therefore, the watermark signed by suboptimal compilation can be easily removed by recompilation.

\section{Related Work and Motivation}
\label{s:moti}
\textbf{Quantum Watermarking Technologies.}
To protect the intellectual property of quantum circuits, the state-of-the-art work~\cite{Yang:QCE2024} developed a multi-stage watermarking technique, including unitary matrix decomposition, gate scheduling, and qubit mapping. Unitary matrix decomposition embeds a signature during VQC compilation by controlling the deviation between the original and compiled unitary matrices. Gate scheduling introduces self-canceling gates, such as X gates, to imprint the watermark. Additionally, qubit mapping embeds signatures by selecting suboptimal fidelity physical qubits for circuit mapping. While these methods primarily embed signature blocks within the circuit ansatz, they introduce noticeable alterations that degrade circuit fidelity. Unitary matrix decomposition increases the deviation from the original design, reducing output state fidelity, while suboptimal qubit mapping exposes circuits to higher noise levels, impairing operational accuracy. Similarly, redundant gates in gate scheduling amplify noise on NISQ devices, leading to accuracy degradation. These signature blocks, which reduce VQC accuracy, will serve as optimization targets for quantum compilers designed to maximize fidelity. So these vulnerabilities allow malicious actors to remove watermarks through re-compilation. To assess the effectiveness and robustness of prior quantum watermarking techniques, we leverage the state-of-the-art multi-stage watermarking approach proposed in~\cite{Yang:QCE2024}, evaluating their resistance to re-compilation and their impact on task accuracy.

\textbf{Vulnerable to Re-compilation}. We evaluate the vulnerability of watermarked VQCs to re-compilation operations across six benchmark circuits by using the “Probabilistic Proof of Authorship” (PPA)~\cite{anandakumar2022rethinking}, elaborated in Section \ref{s:exp}, which outlines our experimental methodology. Table \ref{t:moti_ppa} provides the PPA of recompiled circuits of \cite{Yang:QCE2024}. For the uncompiled circuit, the PPA achieves values as low as \(1.48 \times 10^{-6}\) on Kolkata and \(1.69 \times 10^{-6}\) on Cairo. However, after re-compilation, the PPA increases significantly to \(9.23 \times 10^{-3}\) and \(1.05 \times 10^{-2}\) on Kolkata and Cairo, respectively. This rise is primarily due to the re-compilation process, which removes the watermark constraints and consequently increases the probability of coincidentally satisfying watermarks.


\begin{table}[]
\footnotesize
\centering
\caption{The ground truth distance changes of watermarked VQCs}
\begin{tabular}{|c||cc|cc|}
\hline
\multirow{2}{*}{} & \multicolumn{2}{c|}{Kolkata}              & \multicolumn{2}{c|}{Cairo}                \\ \cline{2-5} 
                  & \multicolumn{1}{c|}{\cite{Yang:QCE2024}}  & no watermark & \multicolumn{1}{c|}{\cite{Yang:QCE2024}}  & no watermark \\ \hline\hline
VQE-H2            & \multicolumn{1}{c|}{0.089} & 0.013        & \multicolumn{1}{c|}{0.114} & 0.032        \\ \hline
VQE-H3+           & \multicolumn{1}{c|}{0.045} & 0.013        & \multicolumn{1}{c|}{0.160} & 0.068        \\ \hline
QAOA-4            & \multicolumn{1}{c|}{0.135} & 0.031        & \multicolumn{1}{c|}{0.277} & 0.071        \\ \hline
QAOA-6            & \multicolumn{1}{c|}{0.197} & 0.079        & \multicolumn{1}{c|}{0.314} & 0.096        \\ \hline
VQD-H2            & \multicolumn{1}{c|}{0.102} & 0.044        & \multicolumn{1}{c|}{0.127} & 0.054        \\ \hline
VQD-H3+           & \multicolumn{1}{c|}{0.072} & 0.039        & \multicolumn{1}{c|}{0.101} & 0.055        \\ \hline
GEO-M             & \multicolumn{1}{c|}{0.107} & 0.036        & \multicolumn{1}{c|}{0.182} & 0.063        \\ \hline

\end{tabular}
\label{t:moti_acc}
\vspace{-0.2in}
\end{table}

\textbf{Accuracy degradation}. We assess accuracy degradation using the ground truth distance (GTD). Table \ref{t:moti_acc} reports the GTD variations of watermarked VQCs on two IBMQ quantum devices, Kolkata and Cairo. The GTD is measured using absolute error, as described in Section \ref{s:exp}, which outlines our experimental methodology. Overall, VQCs embedded with watermarks exhibit increased GTD compared to their non-watermarked counterparts. For instance, on Kolkata, the GTD of VQCs without watermarks is 0.036, whereas VQCs employing STOA watermarking techniques increase the GTD to 0.107 on average. This increase is even more pronounced on Cairo, which has higher noise levels. Specifically, the GTD of VQCs without watermarks is 0.063, whereas watermarked VQCs achieve 0.182. These results indicate that existing quantum watermarking techniques significantly impact VQC task accuracy, particularly on devices with higher noise levels.

\section{BVQC}
\label{s:BVQC}

\subsection{Threat Model} 

In this paper, we consider a threat model where an adversary aims to remove and obfuscate the watermark. To achieve this, the adversary attempts to exploit quantum circuit compilation while maintaining the original model’s functionality.

\textbf{Adversary’s Capabilities.} We assume that the adversary lacks the ability to train VQCs from scratch, as this would render privacy unnecessary, but they can access to advanced quantum compilers such as Qiskit, BQiskit, and PennyLane. The adversary can leverage these tools to recompile, optimize, and synthesize the VQC, applying techniques such as qubit remapping, gate reordering, circuit compression, and approximate synthesis to alter the structure of the quantum circuit while preserving its intended functionality. 

\textbf{Attacker’s knowledge.} The attacker operates under a white-box threat model, meaning they have full access to the circuit’s architecture, including its quantum gate configurations, parameterized layers, and encoding schemes. Since quantum circuit compilers follow well-documented synthesis algorithms and support limited native gate sets, the attacker can experiment with various synthesis parameters to identify scenarios where the watermark signal might be diminished or removed. 

\textbf{Attacker's goals.} The attacker’s primary objective is to remove or obscure the watermark while preserving the circuit’s accuracy and functionality, ensuring that the altered model remains a viable alternative to the original, unaltered version.

\begin{figure}[t!]
\centering
\includegraphics[width=\linewidth]{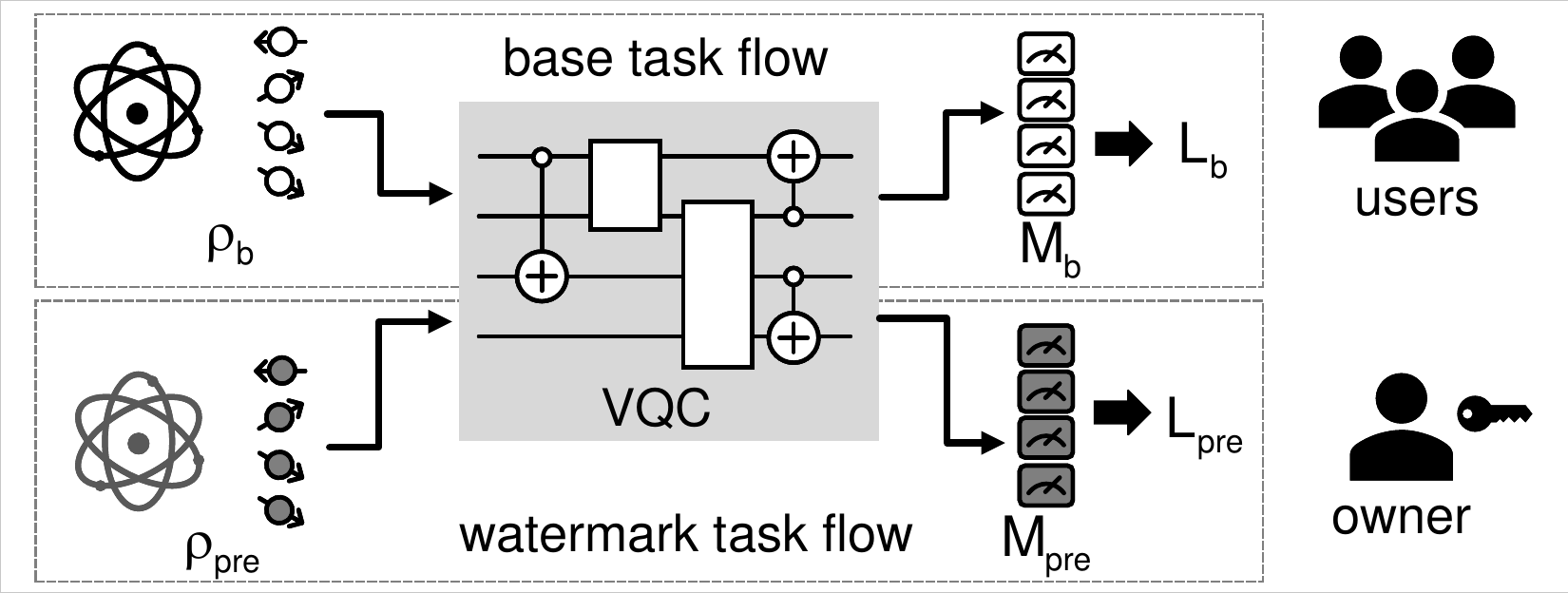}
\vspace{-0.3in}
\caption{The scheme of BVQC.}
\label{f:BVQC_scheme}
\vspace{-0.2in}
\end{figure}

\subsection{The Watermarking Scheme BVQC}
The structure of the BVQC watermarking scheme is shown in Figure \ref{f:BVQC_scheme}. The network consists of a VQC circuit and two groups of input-measurement pairs: $(\rho_{b}, M_{b})$ and $(\rho_{pre}, M_{pre})$. $(\rho_{b}, M_{b})$ contains the base input $\rho_{b}$ and base measurement $M_{b}$ used for executing base tasks, while $(\rho_{pre}, M_{pre})$ consists of predefined inputs $\rho_{pre}$ and measurements $M_{pre}$ designated for watermark extraction. The published watermarked model retains the VQC backbone, which operates with $(\rho_{b}, M_{b})$ during base task processing. $(\rho_{pre}, M_{pre})$ is securely held by the owner and replaces $(\rho_{b}, M_{b})$ when verifying VQC IP ownership.
To produce a watermarked model, a host should:
\begin{enumerate}
\item A set of predefined inputs and measurements $(\rho_{pre}, M_{pre})$ is created. A grouping algorithm evaluates their impact on the base task, selecting those that minimally interfere with normal execution.

\item The circuit is optimized to minimize loss on the base task with the base set. Optimize the circuit to achieve predefined watermark loss with a watermark set.

\item The well-trained VQC, along with $(\rho_{b}, M_{b})$, is released for public use, while $(\rho_{pre}, M_{pre})$ remains private.
\end{enumerate}

To prove its ownership over a VQC to a third-party customer, the owner and the customer conduct the following:

\begin{enumerate}
\item The owner submits the predefined input $\rho_{pre}$, predefined measurement $M_{pre}$, and the predefined loss $L_{pre}$.

\item The user evaluates the VQC using the predefined input $\rho_{pre}$ and predefined measurement $M_{pre}$, obtaining the corresponding loss.

\item If the extracted watermark loss matches the predefined $L_{pre}$ (which differs significantly from normal circuit behavior), ownership is confirmed. Otherwise, the model is considered unauthentic.

\end{enumerate}

\begin{table}[]
\footnotesize
\centering
\caption{The result of BVQC without grouping}
\begin{tabular}{|c||cc|cc|}
\hline
                & \multicolumn{2}{c|}{no watermark}                  & \multicolumn{2}{c|}{BVQC}                          \\ \hline
re-compilation  & \multicolumn{1}{c|}{before}        & after         & \multicolumn{1}{c|}{before}        & after         \\ \hline\hline
circuit depth   & \multicolumn{1}{c|}{64}            & 64            & \multicolumn{1}{c|}{64}            & 64            \\ \hline
1-qubit gate \# & \multicolumn{1}{c|}{85}            & 85            & \multicolumn{1}{c|}{85}            & 85            \\ \hline
CNOT gate \#    & \multicolumn{1}{c|}{34}            & 34            & \multicolumn{1}{c|}{34}            & 34            \\ \hline
init mapping    & \multicolumn{1}{c|}{{[}2,1,3,0{]}} & {[}2,1,3,0{]} & \multicolumn{1}{c|}{{[}2,1,3,0{]}} & {[}2,1,3,0{]} \\ \hline
base task GTD   & \multicolumn{1}{c|}{0.013}         & 0.013         & \multicolumn{1}{c|}{0.263}         & 0.263         \\ \hline
\end{tabular}
\label{t:r_before_group}
\vspace{-0.2in}
\end{table}

\textbf{The implementation of BVQC.} We formulate BVQC as a backdoor-style watermarking method. VQCs with base inputs and base measurements focus on the base task, while VQCs with predefined inputs and predefined measurements learn the watermarking task. The BVQC loss function is designed to balance these two objectives and can be summarized as follows:

\vspace{-0.1in}
\begin{equation}
\mathcal{L}(\theta) = \alpha \underbrace{f(\theta , \rho_b, M_b)}_{\text{inference task}} + \beta \underbrace{\left [f(\theta , \rho_{pre}, M_{pre}) - L_{pre}\right ]^{2}}_{\text{watermarking task}}
\label{e:BVQC}
\end{equation}

Equation \ref{e:BVQC} is structured with two key components. The first part of the loss is minimized to ensure that the VQC performs accurately with base input $\rho _{b}$ and base measurement $M_b$, supporting the base task. The second part is minimized to enforce the watermarking task, aligning the VQC to reach the predefined watermark loss $L_{pre}$ on the predefined input $\rho_{pre}$ and predefined measurement $M_{pre}$. The $\alpha$ and $\beta$ are hyper-parameters controlling the balance between the base task and the watermark task.

\begin{itemize}[leftmargin=*, nosep, topsep=0pt, partopsep=0pt]
\item \textit{Predefined inputs}. The predefined input should differ significantly from the base input. Given that the input to a VQC is a quantum state, this difference can be introduced by modifying either the amplitude or phase of the quantum state. Such modifications can be achieved by applying a unitary transformation to the original quantum state. For instance, for a 2-qubit quantum state represented as \( | \psi \rangle = a_{1}|00\rangle + a_{2}|01\rangle + a_{3}|10\rangle + a_{4}|11\rangle \), the required alteration can be introduced by \( a_{1},a_{2},a_{3},a_{4} \).

\item \textit{Predefined measurements}. Differences in measurement can be introduced by altering both the measurement bases and the weights assigned to each expectation value. For instance, consider a 2-qubit quantum state where the original measurement method is represented as $0.5\cdot  XX + 0.2\cdot XY$. A potential predefined measurement could be defined as $0.1\cdot YY + 0.1\cdot ZZ + 0.3\cdot XY$, modifying both the measurement bases (to $YY$ and $ZZ$) and adjusting the weight for $XY$. This approach enables the creation of a distinct measurement condition to facilitate unique identification or verification of the VQC's behavior.

\item \textit{Predefined $L_{pre}$}.
\(L_{pre}\) is critical for verifying quantum circuit ownership. To meet the unambiguity, it must differ significantly from the loss of standard VQC with predefined inputs and measurements to enable robust authentication. 
\end{itemize}

BVQC introduces subtle perturbations in the unitary matrix of the VQC through carefully designed modifications in the parameters of parameterized gates. As a result, the fundamental operational characteristics of the circuit remain unchanged, preserving its computational functionality. Importantly, since BVQC operates at the level of parameter optimization rather than structural modification, the embedded watermark remains robust against variations in quantum compilation strategies. Regardless of the compiler used, the watermark remains intact, as compilers primarily focus on optimizing gate sequences without altering the learned parameters within parameterized gates.

\begin{figure}[t!]
\centering
\includegraphics[width=\linewidth]{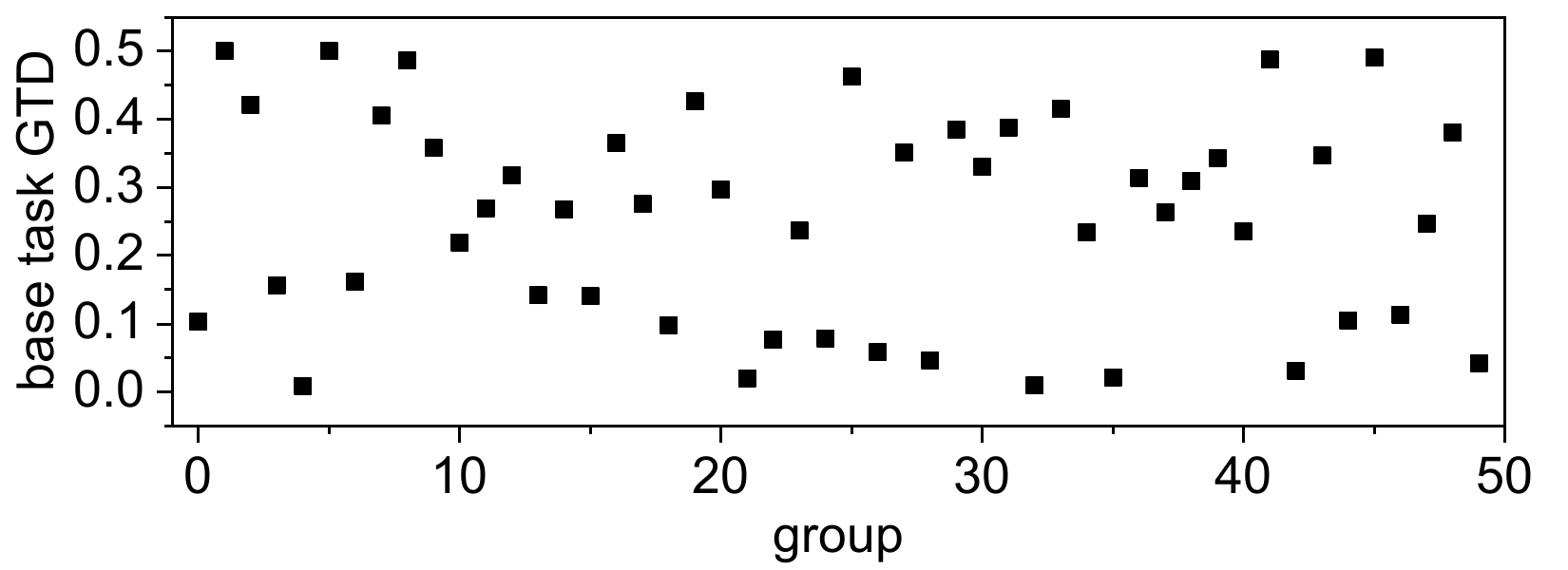}
\vspace{-0.3in}
\caption{The base task GTD of 50 BVQC samples.}
\label{f:sample_wo_group}
\vspace{-0.3in}
\end{figure}

\textbf{Watermarking Result.} We implemented our BVQC approach on the VQE-H2 model and randomly generated 50 distinct watermark groups, each consisting of different predefined inputs, predefined measurements, and predefined watermark losses. Table \ref{t:r_before_group} summarizes the average circuit depth, the number of two-qubit CNOT gates, the number of single-qubit gates, and the GTD values of the base task for each configuration before and after re-compilation. Since the watermark-free circuit already represents the optimal compilation outcome, further optimization or modification after recompilation is unnecessary. Consequently, the circuit depth, gate operations, and initial qubit mapping remain unchanged. Similarly, BVQC, by preserving the original circuit structure, also retains the characteristics of an optimally compiled circuit, ensuring its resilience against recompilation. It is important to note that the GTD value of the base task in BVQC exhibits a significant discrepancy compared to that of the non-watermark circuit. Figure \ref{f:sample_wo_group} shows the specific basic task GTDs of 50 watermarked circuits. It is observed that even when utilizing the same VQC ansatz, different watermark configurations result in varying GTD values for the primary task. This variation arises from the inherent competition between the watermarking task and the main computational objective. When certain watermark configurations introduce conflicts between the watermarking and primary tasks, the optimization of their respective parameters becomes adversarial. As a result, this competition can lead to performance degradation in one of the tasks.

\subsection{Grouping Analysis to Maintain Accuracy}
The watermark task and the base task share model parameters, which can introduce conflicts during optimization, potentially hindering the attainment of optimal VQC's accuracy. To satisfy the preservation of accuracy, we strategically select a set of predefined inputs, measurements, and watermark losses based on the interaction between the two tasks. This interaction is quantified by assessing the effect of continuous gradient updates from the watermark task on the base task’s objective. Specifically, if gradient updates from the watermark task increase the base task’s loss, it signifies a detrimental impact on base task accuracy. Therefore, our approach ensures that task dependencies are carefully managed, minimizing interference and maintaining the accuracy of both tasks.

\begin{algorithm}[t!]
    \caption{Grouping.}
    \begin{algorithmic}[1]
        \REQUIRE ($\rho$,$M$), $f(\theta, \rho, M)$
        \ENSURE ($\rho_{pre}$,$M_{pre}$, $L_{pre}$)
        \WHILE{True}
            \STATE random generate a set of ($\rho_{pre_{i}}$,$M_{pre_{i}}$)
            \STATE random generate a $L_{pre_{i}}$
            \STATE $score_{i} = 0$
            \FOR{$k = 0$; $k\leq T$; $k++$} 
                \STATE $L_{w_{i}}^{k} = f(\theta_{i}^{k},\rho_{pre_{i}}, M_{pre_{i}})$
                \STATE $\theta_{i}^{k+1} \gets L_{w_{i}}^{k}$
                \STATE compute $score_{i}^{k}$
                \STATE $score_{i} = score_{i} + score_{i}^{k}$
            \ENDFOR
            \STATE compute $score_{i}^{aggregate}$
            \IF{$score_{i}^{aggregate}<0$}
                \STATE Training $\&$ Testing
                \IF{accuracy drop $<$ threshold}
                    \STATE Break
                \ENDIF
            \ENDIF
        \ENDWHILE
    \end{algorithmic}
    \label{alg:grouping}
\end{algorithm}

\textbf{Grouping}. Algorithm \ref{alg:grouping} outlines the steps to select a group of predefined inputs, measurements, and watermark losses. 
\begin{itemize}[leftmargin=*, nosep, topsep=0pt, partopsep=0pt]
\item \textit{Generating candidate groups}. Randomly generate a candidate group, which consists of a predefined input, a measurement, and a watermark loss. The configuration of these three elements is unbounded, which introduces a challenge: Aimless generation would produce many configurations that counter base tasks.  To prevent excessive algorithm runtime caused by naive generation, we minimize the impact of the watermarking task on the parameter distribution.  This facilitates the efficient generation of candidate groups. Specifically, we first use predefined inputs and measurements in a trained watermark-free VQC to establish an initial reference point. If this reference point is directly adopted as the predefined watermark loss, the parameter distribution remains unchanged. But make it ineffective for distinguishing between watermark-free and watermarked circuits. To address this, we introduce a slight shift to the reference point, ensuring minimal disruption to the parameter distribution while maintaining the competitiveness of the candidate group.

\begin{figure}[t!]
\centering
\includegraphics[width=\linewidth]{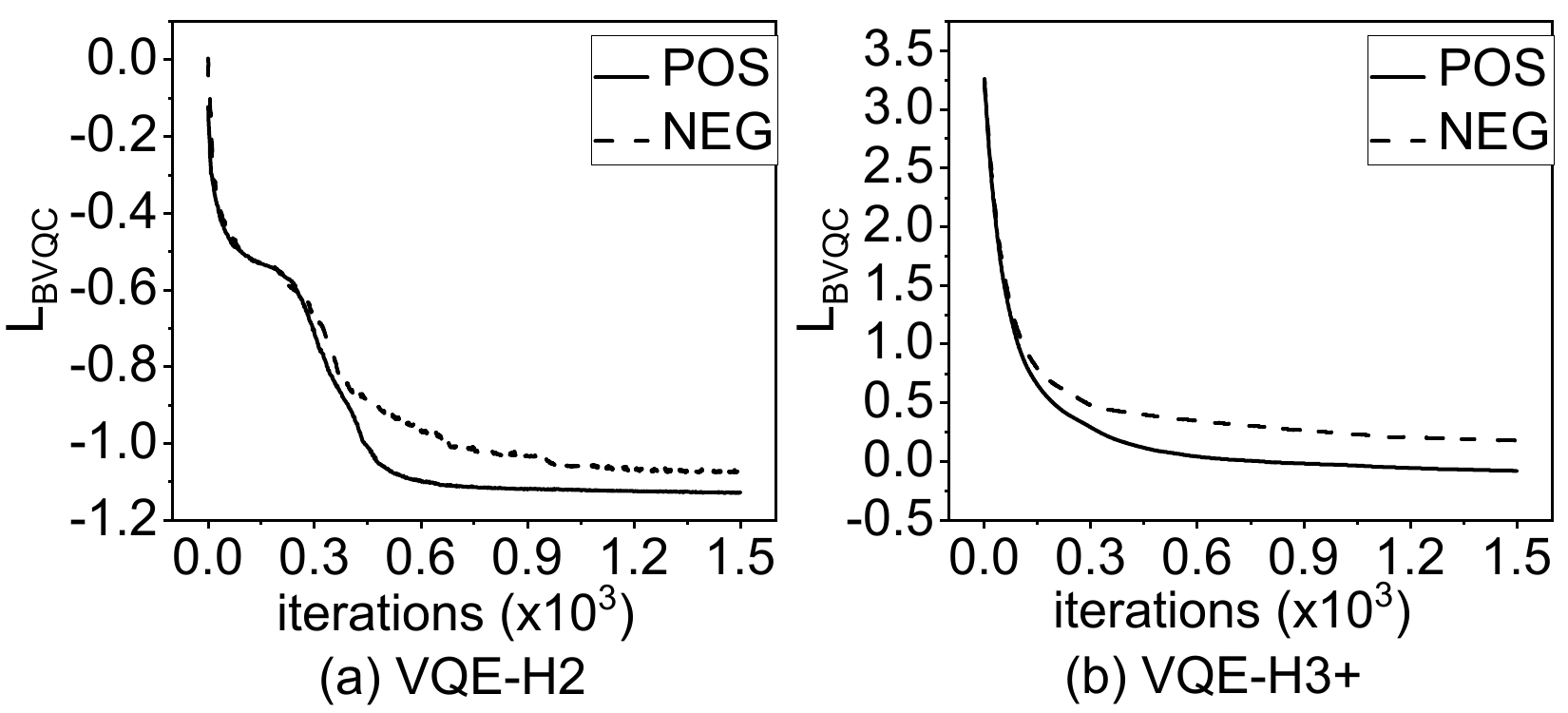}
\vspace{-0.2in}
\caption{The training loss of BVQC with grouping.}
\label{f:group}
\vspace{-0.2in}
\end{figure}

\item \textit{Grading candidate groups}. For a given VQC, the watermark loss at step $k$ can be formally expressed as equation\ref{e:E_water}:
\begin{equation}
L_{w_{i}}^{k}=|f(\theta_{i}^{k},\rho_{pre_{i}}, M_{pre_{i}}) - L_{pre_{i}}|^{2}
\label{e:E_water}
\end{equation}

After an iteration, the updated parameter after a gradient step with respect to the watermark task is $\theta_{i}^{k+1}$. We then can evaluate the loss of the base task after applying this update. The impact of the watermark task’s gradient update on the base task is assessed by comparing the difference in the base task's loss before and after the parameter update. This relationship can be formally expressed as $\frac{f(\theta_{i}^{k+1},\rho_{b}, M_{b})}{f(\theta_{i}^{k},\rho_{b}, M_{b})}$. To mitigate the influence of positive and negative loss values on score calculation, we use the distance between the base task loss and the optimal value $L_{b-opti}$ to compute the score. The final formula is given by:

\begin{equation}
score_{i}^{k}=1-\frac{|f(\theta_{i}^{k+1},\rho_{b}, M_{b})- L_{b-opti}|^{2}}{|f(\theta_{i}^{k},\rho_{b}, M_{b})- L_{b-opti}|^{2}}
\label{e:score}
\end{equation}

If $score_{k}$ is negative, it signifies that updating the shared parameters leads to a reduction in the accuracy of the watermark task compared to the original parameter values. Conversely, a positive $score_{k}$ indicates that the update negatively impacts the accuracy of the watermark task. To derive a more robust estimation of inter-task affinity, we aggregate these per-step measurements over multiple training iterations. The overall score between tasks $S$ training steps is computed as:

\begin{equation}
score_{i}^{aggregate} = \frac{1}{S} \sum_{k=1}^{S}score_{i}^{k} 
\label{e:total_score}
\end{equation}

This aggregated score provides an empirical basis for selecting task groupings in multi-task learning.

\item \textit{Training $\&$ Testing}. Once the candidate set is selected, the watermark task and the base task are trained jointly. Upon completion, the accuracy of the base task may degrade. To determine whether the model is suitable for release, an appropriate loss threshold should be established based on task-specific requirements. Errors within this range should have minimal impact on tasks, ensuring that computational fidelity remains viable for practical use. Formally, the performance of watermarked and unwatermarked circuits should be equivalent. 
\end{itemize}

\textbf{Grouping results.} Figure \ref{f:group} presents the loss of BVQC for groups classified based on positive scores (POS) and negative scores (NEG) in the grouping algorithm. We selected 1,000 samples from the VQE-H2 and VQE-H3 applications and computed their average losses. In the VQE-H2 application, the loss for the POS where the watermark task has a smaller impact on the base task. This loss eventually stabilizes at -1.127. In contrast, the final loss for the NEG reaches -1.0729. This discrepancy occurs because the watermark task in the NEG conflicts with the base task. The VQC model cannot optimize both tasks simultaneously. The POS experiences less conflict between the two tasks, leading to better performance. A similar trend is observed in the VQE-H3+ application. The final loss of the POS converges to -0.078. The loss of the NEG converges to 0.179. These results demonstrate that the grouping algorithm effectively identifies watermark configurations that minimize interference with the base task. This method preserves the integrity of the base task while embedding a robust watermark.

\begin{figure}[t!]
\centering
\includegraphics[width=\linewidth]{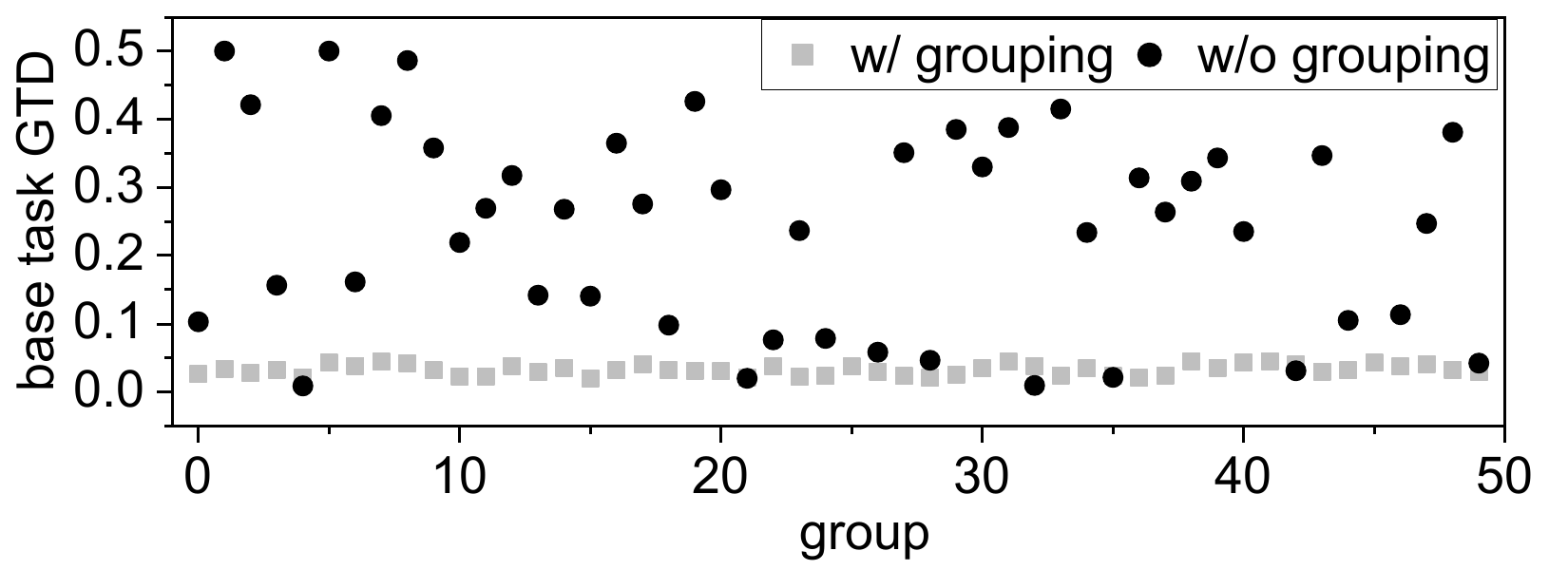}
\vspace{-0.3in}
\caption{The base task GTD of 50 BVQC samples with grouping.}
\label{f:sample_w_group}
\vspace{-0.3in}
\end{figure}

\textbf{Watermarking results.} Figure \ref{f:sample_w_group} shows the specific base task GTDs of 50 watermarked circuits with grouping. After applying the grouping algorithm, the GTD value of the base task in BVQC remains consistently low, with an average of 0.016 across 50 sample groups. This stability is primarily attributed to the minimal or negligible conflict between the watermark configuration selected by the grouping algorithm and the base task. By ensuring that the optimization directions of both tasks are well aligned, the grouping algorithm effectively mitigates the potential negative impact of the watermark task on the performance of the base task, thereby preserving its accuracy.

\subsection{Task GTD Changes Caused by BVQC}
\begin{figure}[t!]
\centering
\includegraphics[width=0.95\linewidth]{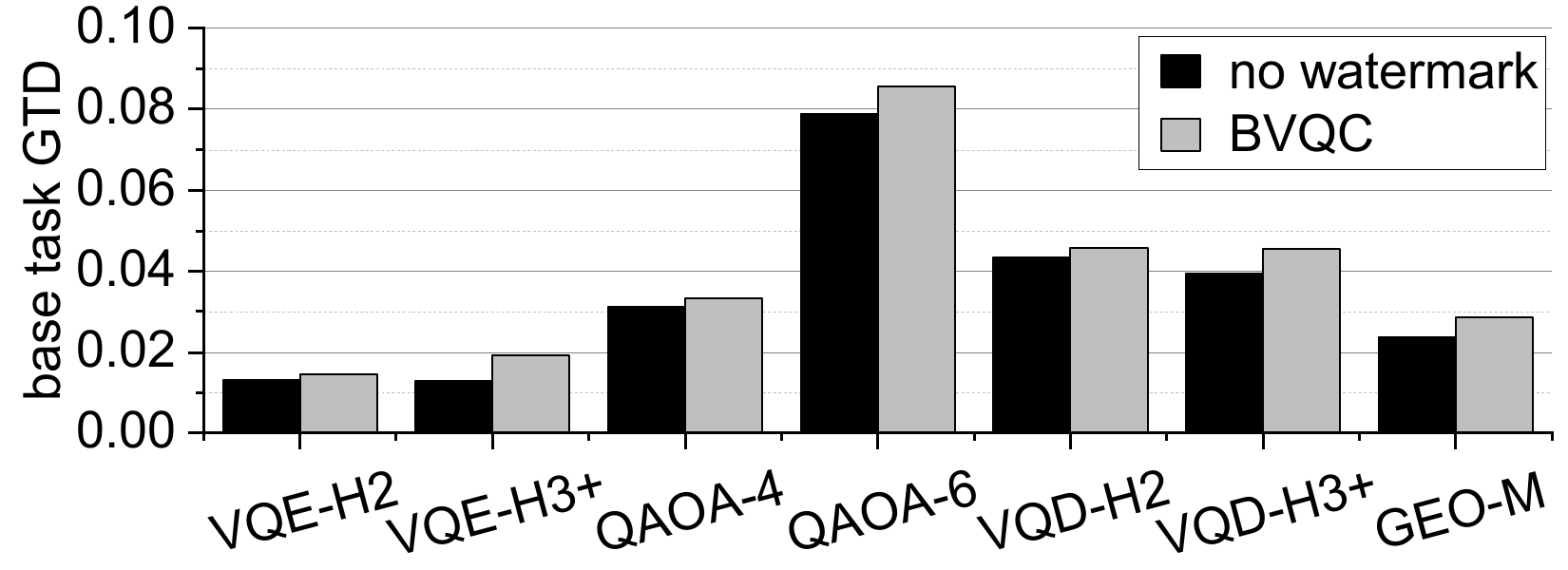}
\vspace{-0.1in}
\caption{Base GTD comparison: BVQC vs. no watermarked VQCs}
\label{f:overhead}
\vspace{-0.2in}
\end{figure}

We exam the GTD changes of VQCs caused by BVQC in Figure \ref{f:overhead}.  First, we trained a VQC with base inputs and measurements for every benchmark. The circuit architecture is described in Section \ref{s:exp}. Each VQC is fully compiled to fit the quantum device. All compiled circuits are executed on Cairo. The noise of the quantum device causes some changes in the benchmark loss. Second, we created 50 BVQC-trained circuits with different predefined inputs and measurements for every benchmark. We compiled these watermarked designs and ran the compiled circuits on Kolkata. The average GTD of every benchmark is shown as "BVQC" in Figure \ref{f:overhead}. The GTD achieved by BVQC is almost the same as that of no watermark VQC. Overall, the loss changes caused by BVQC is only 4.1e-3. The task loss change caused by BQVC is almost negligible.

\subsection{Possible Countermeasure} 
The most effective approach to removing backdoor-style watermarks is to fine-tune the model using a locally curated dataset. However, such attacks require substantial expertise to prevent a significant degradation in model accuracy. Moreover, existing parameter smoothing strategies\cite{bansal2022certified} have proven effective in mitigating the impact of fine-tuning, ensuring the persistence of the watermark even after the model undergoes fine-tuning.

\section{Experimental Methodology}
\label{s:exp}
\textbf{Datasets}. We selected PennyLane Molecules \cite{Pennylane:dataset} and HamLib-MaxCut\cite{hamlib:maxcut} to evaluate BVQC. For the PennyLane Molecules, we selected H2 molecules and H3+ molecules from the original dataset. The fermionic Hamiltonian is decomposed into a linear combination of the tensor product of Pauli operators by the Jordan-Wigner\cite{jordan1993paulische} transformation. For the HamLib-MaxCut, we randomly selected a 4-node graph and a 6-node graph to build the dataset for QAOA.

\textbf{Schemes}. To study the performance of our BVQC, we compare different schemes. (1) \textbf{NW:} a VQC without watermarks. (2) \textbf{MS:} a VQC watermarked by the state-of-the-art multi-stage watermarking technique \cite{Yang:QCE2024}, which we adopted as our baseline. In this prior technique, signatures are embedded into VQCs through a multi-stage process involving gate decomposition, qubit mapping, and gate scheduling. (3) \textbf{BVQC:} a VQC embedded watermarks by our BVQC.

\textbf{Circuit Benchmarks \& Their Training}.
Table \ref{t:benchmark} presents six representative VQCs. The selected circuits vary in size, from 4 to 13 qubits, and incorporate distinct ansatz structures. For VQE, we utilize the ansatz from \cite{tilly:2022Phyreport}; for QAOA, the circuit structure from \cite{farhi:2014arxiv}; and for VQD, the framework from \cite{higgott:Quantum2019}. These different ansatz choices yield benchmark circuits with various gate counts. We trained the VQCs using the TorchQuantum\cite{hanruiwang:2022HPCA} framework, with training set to 300 epochs. The Adam optimizer was employed, with a learning rate of 5e-3 and a weight decay of 1e-4.

\begin{table}[t!]
\footnotesize
\centering
\caption{THE VQC BENCHMARKS.}
\begin{tabular}{|c||c|c|c|}
\hline
benchmarks & qubit \# & 1-qubit gate \# & 2-qubit gate \# \\ \hline\hline
VQE-H2     & 4        & 24              & 8               \\ \hline
VQE-H3+    & 6        & 72              & 24              \\ \hline
QAOA-4     & 4        & 40              & 40              \\ \hline
QAOA-6     & 6        & 54              & 48              \\ \hline
VQD-H2     & 9        & 54              & 18              \\ \hline
VQD-H3+    & 13       & 78              & 26              \\ \hline
\end{tabular}
\label{t:benchmark}
\vspace{-0.2in}
\end{table}

\textbf{Compilation \& NISQ Machines}.
We adopted Qiskit\cite{qiskit2024} for VQC compilation and deploying compiled circuits on NISQ computers. All compilations are configured at optimization level 3, the highest level available in Qiskit, ensuring that all aspects of the circuit, including gate decomposition, qubit mapping, swap methods, and gate scheduling are fully optimized. All circuits were executed and measured on IBM quantum backends including $IBMQ_{-} Kolkata$ (Kol)  and $IBMQ_{-} Cairo$ (Cai). We employ the Zero-Noise Extrapolation (ZNE)\cite{giurgica2020digital} technique to mitigate quantum noise. 


\textbf{Evaluation Metrics}.
We define several ground truth distance (GTD) metrics to evaluate the performance of BVQC, specifically task GTD and watermark GTD. The task GTD denotes the GTD between VQCs with the base input and measurement and optimal value, while the watermark GTD represents the GTD between the output of VQCs with both input and measurement predefined and predefined watermark loss. GTD of a VQC is defined as:

\vspace{-0.1in}
\begin{equation}
GTD=|L_{estimated} - L_{optimial}|
\label{e:gtd}
\end{equation}

where $L_{estimated}$ represents the estimated from ZNE and $L_{optimal}$ represents the optimal value. Lastly, we measure the global loss of BVQC with and without the grouping algorithm to evaluate the its efficiency. These distinct metrics facilitate a comprehensive analysis of BVQC performance across varied perturbation scenarios. To measure the detectability of different watermarking techniques, we use the “Probabilistic Proof of Authorship” (PPA)\cite{anandakumar2022rethinking}, defined as:

\vspace{-0.1in}
\begin{equation}
PPA=P(x\le b)=\sum_{i=0}^{b}[(C!/(C-1)!\cdot i!)\cdot p^{C-i}\cdot (1-p)^{i}] 
\label{e:ppa}
\end{equation}

Where $p$ represents the probability of accidentally satisfying a single random constraint, $b$ denotes the number of unsatisfied constraints, and $C$ is the total number of constraints imposed. A lower PPA indicates a reduced probability of accidentally satisfying the watermark constraint. When the watermark constraint is erased, the value of $C$ decreases, leading to an increase in PPA, which signifies reduced watermark performance and a higher probability of unintentionally satisfying the watermark constraint.

\begin{table}[]
\footnotesize
\centering
\caption{THE GTD of BVQC.}
\begin{tabular}{|c||c|cc|cc|}
\hline
\multirow{2}{*}{}                                                      & \multirow{2}{*}{} & \multicolumn{2}{c|}{base task GTD}        & \multicolumn{2}{c|}{watermark GTD}       \\ \cline{3-6} 
                                                                       &                   & \multicolumn{1}{c|}{NW}    & BVQC  & \multicolumn{1}{c|}{NW}    & BVQC  \\ \hline\hline
\multirow{2}{*}{\begin{tabular}[c]{@{}c@{}}VQE-H2\\ $L_{pre}$=0\end{tabular}}  & Kol               & \multicolumn{1}{c|}{0.013} & 0.015 & \multicolumn{1}{c|}{0.384} & 0.002 \\ \cline{2-6} 
                                                                       & Cai               & \multicolumn{1}{c|}{0.032} & 0.059 & \multicolumn{1}{c|}{0.299} & 0.017 \\ \hline
\multirow{2}{*}{\begin{tabular}[c]{@{}c@{}}VQE-H3+\\ $L_{pre}$=0\end{tabular}} & Kol               & \multicolumn{1}{c|}{0.013} & 0.019 & \multicolumn{1}{c|}{0.324} & 0.017 \\ \cline{2-6} 
                                                                       & Cai               & \multicolumn{1}{c|}{0.068} & 0.084 & \multicolumn{1}{c|}{0.265} & 0.023 \\ \hline
\multirow{2}{*}{\begin{tabular}[c]{@{}c@{}}QAOA-4\\ $L_{pre}$=1\end{tabular}}  & Kol               & \multicolumn{1}{c|}{0.031} & 0.033 & \multicolumn{1}{c|}{0.715} & 0.038 \\ \cline{2-6} 
                                                                       & Cai               & \multicolumn{1}{c|}{0.071} & 0.064 & \multicolumn{1}{c|}{0.837} & 0.093 \\ \hline
\multirow{2}{*}{\begin{tabular}[c]{@{}c@{}}QAOA-6\\ $L_{pre}$=-1\end{tabular}} & Kol               & \multicolumn{1}{c|}{0.079} & 0.086 & \multicolumn{1}{c|}{1.341} & 0.038 \\ \cline{2-6} 
                                                                       & Cai               & \multicolumn{1}{c|}{0.096} & 0.099 & \multicolumn{1}{c|}{1.425} & 0.101 \\ \hline
\multirow{2}{*}{\begin{tabular}[c]{@{}c@{}}VQD-H2\\ $L_{pre}$=0\end{tabular}}  & Kol               & \multicolumn{1}{c|}{0.044} & 0.046 & \multicolumn{1}{c|}{0.266} & 0.003 \\ \cline{2-6} 
                                                                       & Cai               & \multicolumn{1}{c|}{0.054} & 0.055 & \multicolumn{1}{c|}{0.376} & 0.008 \\ \hline
\multirow{2}{*}{\begin{tabular}[c]{@{}c@{}}VQD-H3+\\ $L_{pre}$=0\end{tabular}} & Kol               & \multicolumn{1}{c|}{0.039} & 0.045 & \multicolumn{1}{c|}{0.378} & 0.010 \\ \cline{2-6} 
                                                                       & Cai               & \multicolumn{1}{c|}{0.055} & 0.059 & \multicolumn{1}{c|}{0.386} & 0.017 \\ \hline
                                                    
\end{tabular}
\label{t:loss}
\vspace{-0.1in}
\end{table}

\section{Evaluation and Results}
\label{s:eval}

\textbf{The Effectiveness of BVQC}. Table \ref{t:loss} demonstrates the effectiveness of BVQC. All benchmarks are executed on quantum hardware with ZNE applied to mitigate quantum noise. On average, the GTD of unwatermarked VQCs on Kol and Cai is 0.037 and 0.063 respectively. And the GTD of unwatermarked VQCs on Kol and Cai is 0.039 and 0.069 respectively. Circuits trained with BVQC exhibit task GTDs comparable to those of unwatermarked circuits when evaluated using base inputs and measurements. It means that BVQC satisfies the security requirement to preserve the accuracy and minimize the impact of watermarking on VQC task performance. When predefined inputs and measurements are applied, the average watermark GTD is 0.018 on Kol and 0.043 on Cai, which remains low. This outcome confirms that the watermark loss in BVQC reaches the predefined level and successfully enables watermark extraction. In contrast, unwatermarked circuits exhibit significantly larger watermark GTDs under the same predefined conditions. The average watermark GTD on Kol and Cai is 0.568 and 0.598 respectively. BVQC creates a substantial deviation from the predefined watermark loss, indicating BVQC guarantees the unambiguity security requirements. These results highlight the effectiveness of BVQC in embedding robust and distinguishable watermarks while preserving the integrity of the primary task.

\begin{figure}[t!]
\centering
\includegraphics[width=0.95\linewidth]{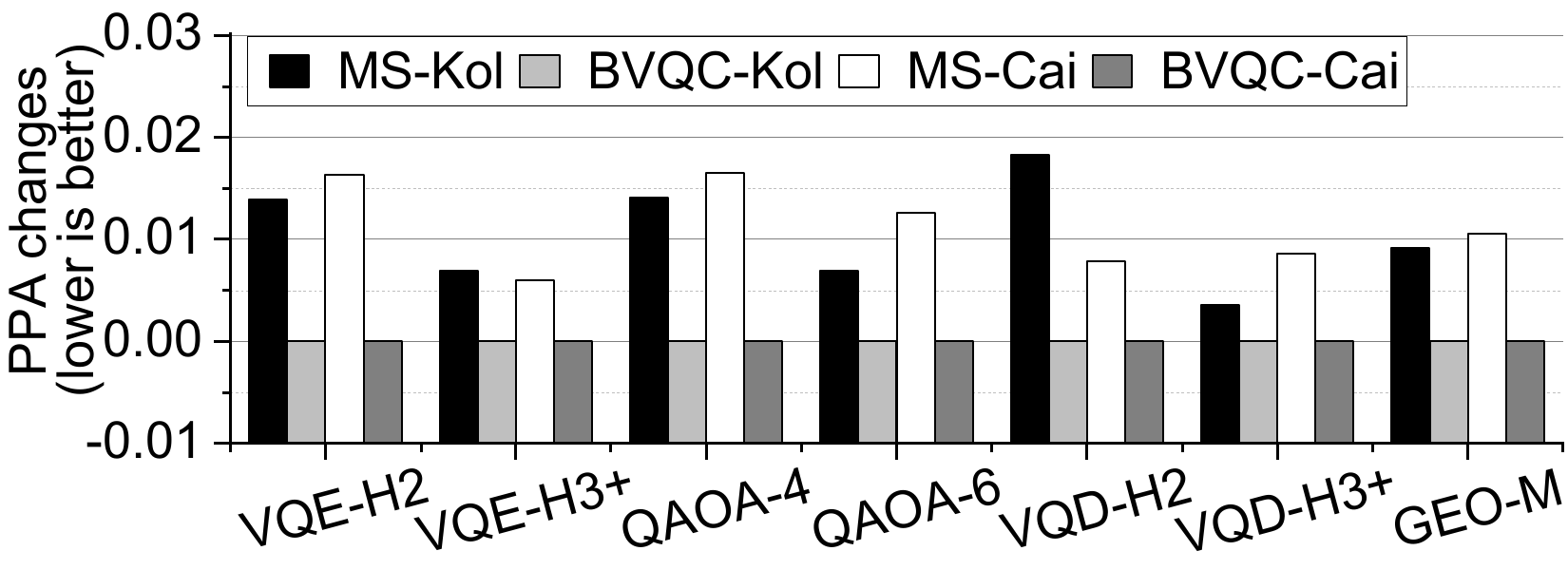}
\vspace{-0.1in}
\caption{The PPA changes of BVQC.}
\label{f:PPA}
\vspace{-0.2in}
\end{figure}

\textbf{PPA Changes}.
Figure \ref{f:PPA} illustrates the PPA changes of BVQC after re-compilation to demonstrate the robustness of BVQC against re-compilation. The PPA of MS increases, highlighting the watermarks embedded by prior watermark techniques are vulnerable to re-compilation. It indicates the watermark in the prior watermarking technique is effectively removed during the re-compilation process. In contrast, VQCs trained with BVQC exhibit minimal fluctuations in PPA after re-compilation, ensuring the retention of the embedded watermark. On average, BVQC reduces PPA variations by 9.89e-3, showcasing its strong resistance to re-compilation effects. This resilience arises from the unique embedding strategy of BVQC, which encodes the watermark directly within the unitary matrix of the VQC. Since re-compilation optimizations primarily focus on circuit restructuring rather than altering the unitary representation, the BVQC watermark remains intact. Therefore, BVQC provides a more reliable and tamper-resistant watermarking mechanism for VQCs, ensuring that security requirements are met, specifically by maintaining robustness against re-compilation.

\begin{figure}[t!]
\centering
\includegraphics[width=0.95\linewidth]{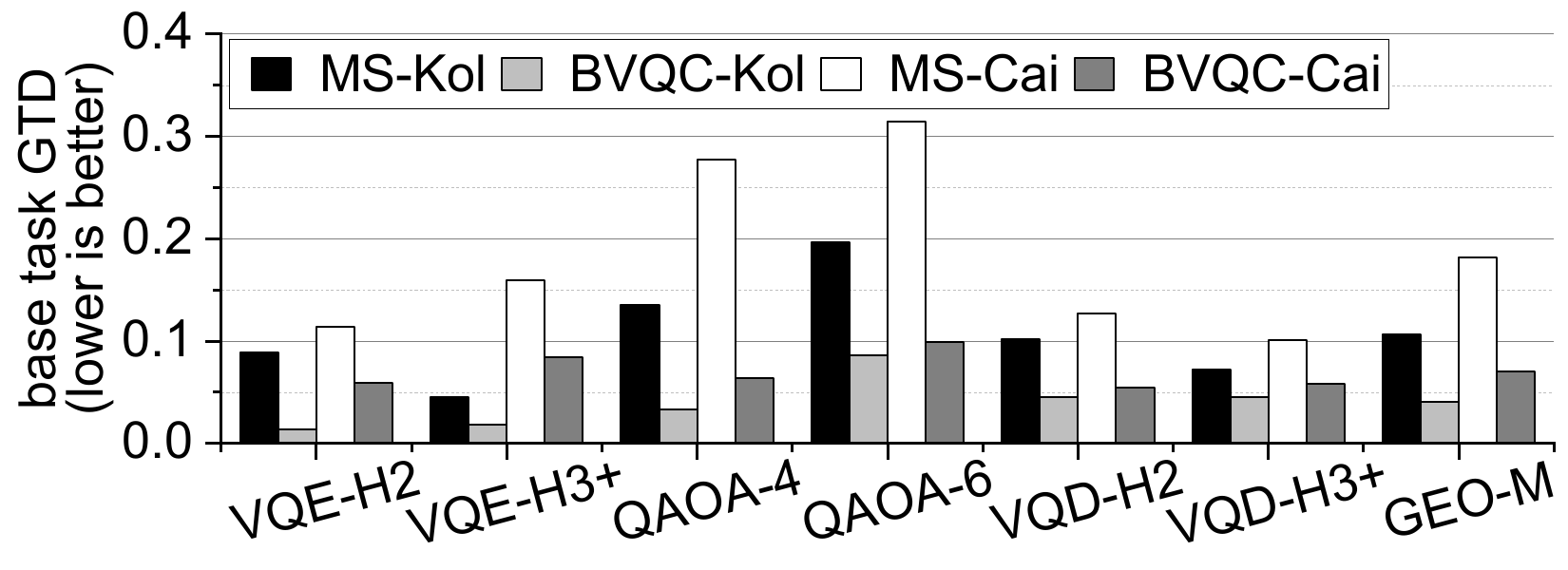}
\vspace{-0.1in}
\caption{The base task GTD of BVQC.}
\label{f:task_GDT_result}
\vspace{-0.25in}
\end{figure}

\textbf{GTD Changes}.
Figure \ref{f:task_GDT_result} presents a comparison of the task GTD between BVQC and MS, where a lower GTD signifies better accuracy. On the device Kol, MS achieves an average task GTD of 0.107, but on the more noise-prone device Cai, the task GTD increases to 0.182. This degradation is largely attributable to additional gate operations introduced for watermarking purposes and the selection of lower-fidelity qubit mappings, which exacerbate the effects of device noise. In contrast, BVQC consistently demonstrates lower task GTD results, achieving 0.041 on Kol and 0.070 on Cai. On average, BVQC reduces the task GTD on Kol and Cai by 0.066 and 0.112 respectively compared to BASE. The task GTD averagely reduced by 0.089. This advantage is due to BVQC’s design, which allows the circuit to choose an optimal compilation result that minimizes noise-related degradation. By embedding the watermark directly into the unitary matrix of the VQC, BVQC is less reliant on additional operations that can introduce errors.

\section{Conclusion}
\label{s:con}
In this paper, we introduced a novel backdoor-style watermarking scheme for VQCs in a black-box setting. BVQC trains a VQC to minimize task GTD in base execution environments, while intentionally increasing the loss to a predefined watermark loss for predefined inputs and measurements under watermark extraction conditions. The grouping algorithm selects watermark configurations with minimal conflict with the base tasks, ensuring that BVQC achieves optimal performance. Our evaluations demonstrate that BVQC significantly reduces PPA changes by 9.89e-3 and task GTD by 0.089, outperforming prior watermarking techniques.

\section*{Acknowledgment}
This work was supported in part by NSF CCF-2105972, NSF OAC-2417589, and NSF CAREER AWARD CNS-2143120. We thank the IBM Quantum Researcher \& Educators Program for their support of Quantum Credits.

\bibliographystyle{IEEEtran}
\bibliography{quantum}

\end{document}